\documentclass[aps,prd,twocolumn,amsmath,floatfix,longbibliography]{revtex4-1}
\usepackage{lineno,isotope,mathrsfs}
\modulolinenumbers[5]
\usepackage{amsmath,bbold,amssymb,epsfig,bm,mathrsfs,feynmp}
\usepackage{color,slashed,nicefrac,exscale,multirow,soul}
\usepackage{times,txfonts,xcolor,graphicx,gensymb,tikz}
\usepackage[normalem]{ulem}
\usepackage[breaklinks,colorlinks,citecolor=blue]{hyperref}
\definecolor{red}{rgb}{0.8,0,0}
\definecolor{violet}{rgb}{0.4,0,0.4}
\definecolor{green}{rgb}{0,0.5,0.0}
\definecolor{navy}{rgb}{0.0,0.0,0.6}
\definecolor{orange}{rgb}{0.8,0.2,0.0}

\newcommand{\bea}{\begin{eqnarray}}
\newcommand{\eea}{\end{eqnarray}}
\newcommand{\ep}{\varepsilon}
\begin{document}
\title{Relativistic hybrid stars with sequential first-order phase
transitions \\ and heavy-baryon envelopes}
\author{Jia Jie Li}
\email{jiajieli@itp.uni-frankfurt.de}
\affiliation{Institute for Theoretical Physics,
J. W. Goethe University,
D-60438 Frankfurt am Main, Germany}
\author{Armen Sedrakian}
\email{sedrakian@fias.uni-frankfurt.de}
\affiliation{Frankfurt Institute for Advanced Studies,
D-60438 Frankfurt am Main, Germany}
\affiliation{Institute of Theoretical Physics,
University of Wroclaw, 50-204 Wroclaw, Poland}
\author{Mark Alford}
\email{alford@physics.wustl.edu}
\affiliation{Department of Physics, Washington University,
St.~Louis, Missouri 63130, USA}
\begin{abstract}
We compute the mass, radius and tidal deformability of stars
containing phase transitions from hadronic to quark phase(s).
These quantities are computed for three types of hadronic envelopes:
purely nuclear, hyperonic, and $\Delta$-resonance--hyperon admixed
matter. We consider either a single first-order phase transition to
a quark phase with a maximally stiff equation of state (EOS) or two
sequential first-order phase transitions mimicking a transition from
hadronic (H) to a quark matter phase followed by a second phase
transition to another quark phase. Such a construct emulates the
results of the computations of the EOS which include 2SC and CFL
color superconducting phases at low and high density. We explore
the parameter space which produces low-mass twin and triplet
configurations where equal-mass stars have substantially different
radii and tidal deformabilities. We demonstrate that while for
purely hadronic stiff EOS the obtained maximum mass is inconsistent
with the upper limit on this quantity placed by GW170817, the
inclusion of the hyperonic and $\Delta$-resonance degrees of
freedom, as well as the deconfinement phase transition at
sufficiently low density, produces a configuration of stars consistent
with this limit. The obtained hybrid star configurations are in the
mass range relevant for the interpretation of the GW170817 event.
We compare our results for the tidal deformability with the limits
inferred from GW170817 showing that the onset of non-nucleonic
phases, such as $\Delta$-resonance--hyperon admixed phase and/or the
quark phase(s), are favored by these data if the nuclear EOS is stiff.
Also, we show that low-mass twins and especially triplets proliferate
the number of combinations of possible types of stars that can
undergo a merger event, the maximal number being six in the case
of triplets. The prospects for uncovering the first-order phase
transition(s) to and in quark matter via measurements of tidal
deformabilities in merger events are discussed.
\end{abstract}
\date{\today}
\maketitle
%
\section{Introduction}
\label{sec:Intro}

The first multimessenger observations of gravitational waves (GW)
by the LIGO-Virgo Collaboration from the binary neutron star merger
event GW170817~\cite{LIGO_Virgo2017c,LIGO_Virgo2017a,LIGO_Virgo2017b}
have provided important new constraints on the equation of state (EOS)
of dense matter through the measurement of the tidal deformabilities
and masses of neutron stars involved. The combination of these
results with the recent measurement of the highest mass of a millisecond
pulsar, $M = 2.14^{+0.10}_{-0.09}M_\odot$ (within a 68.3\% credibility
interval) for pulsar PSR J0740+6620~\cite{Cromartie2019}, provides the
most stringent astrophysical constraints to date on the properties of
ultra-dense matter. To be consistent with these data, the EOS of the
dense matter must be moderately soft at intermediate densities to
allow for relatively small tidal deformabilities but it must be hard
enough at high densities to allow for two-solar-mass neutron stars.

Constraints on the EOS of dense matter were placed also using
the pulse-profile modeling of the isolated 205.53 Hz millisecond
pulsar PSR J0030+0451 observed in x-rays by the Neutron Star
Interior Composition Explorer (NICER) experiment. The two
independent analysis, which used different emitting spot models,
constrained the mass and equatorial radius of this object
within a $68.3\%$ credibility interval to
$1.34^{+0.15}_{-0.16} M_{\odot}$ and $12.71^{+1.14}_{-1.19}$ km
\cite{Riley2019} and $1.44^{+0.15}_{-0.14} M_{\odot}$ and
$13.02^{+1.24}_{^-1.06}$ km~\cite{Miller2019}.

The actual composition of matter in the interior of neutron stars
remains unclear. One possibility is a deconfinement phase transition
from bound hadronic states to liberated quark states (see, for
instance, Refs.~\cite{Benic2015,Kaltenborn2017,Fischer2018,Bauswein2019}
and references therein). The effect of a {\it strong first-order} phase
transition on tidal deformabilities of the neutron star in the context
of the GW170817 event has been explored by a number of groups~\cite{Annala2018,
Paschalidis2018,Nandi2018,Most2018,Tews2018,Burgio2018,Alvarez-Castillo2019,
Christian2019,HanSophia2019,Montana2019,Sieniawska2019,Essick2019} showing
that the hadron-quark phase transition at low enough density, $\rho \simeq 2$-$3\rho_0$,
where $\rho_0$ is the nuclear saturation density, relaxes the tension
between the inferred tidal deformabilities and those predicted by
purely hadronic models without a phase transition. Of particular
interest in this context is the emergence of {\it twin} stars, where
purely nucleonic and hybrid stars have the same masses, but different
radii~\cite{Paschalidis2018}.

The onset in dense matter of heavy baryonic states, such as hyperons
and $\Delta$-resonances, was explored in
Refs.~\cite{Lijj2019a,Lijj2019b}, taking into account the data from
GW170817. These works showed that the EOS, in this case, can be hard
enough to support two-solar-mass neutron stars and that
$\Delta$-resonances allow for intermediate-density-range softening of
the EOS, which is in line with the low values of deformabilities
inferred for $M\simeq 1.4M_{\odot}$ stars. However, the matter
containing heavy baryonic states may undergo a phase transition to
quark matter at densities relevant for compact stars. Therefore,
phenomenologically, it is of great interest to study a sequence of
phase transitions, where the purely nucleonic matter undergoes a phase
transition to matter with heavy baryon states (hyperons and
$\Delta$-resonances) at a density somewhat above the nuclear
saturation density which is followed by deconfinement of hadrons to
quarks at some higher density. Below, for the sake of brevity, the
conglomerate of hadronic phases of a hybrid star, which surrounds its
quark core is referred as {\it hadronic envelope.}

Thus, the first motivation of this work is to study the tidal
deformabilities of a neutron star with such a sequence of phase
transitions. The second motivation of this work is to compute the
tidal deformabilities of hybrid stars with sequential first-order
phase transitions~\cite{Alford2017}. For example, this scenario can
be realized if with increasing density the hadronic matter makes a
transition to an intermediate-density quark phase ``Q1''
[e.g., the two-flavor color-superconducting (2SC) quark phase~\cite{Bailin1984}],
followed by a second transition to a higher-density quark phase ``Q2''
[e.g., the three-flavor color-flavor-locked (CFL) quark
phase~\cite{Alford1999}]. Although we are concerned with the
phenomenological description of such phase transition, Nambu-Jona-Lasinio
(NJL) model-based computations~\cite{Bonanno2012} confirm the possibility
of such an outcome if a repulsive vector interaction is added to the
standard NJL Lagrangian.

In this work we explore the regime where the phase transitions take place
at low enough density so that low-mass stars with $M\simeq 1.4 M_{\odot}$
may contain quark matter. It has been shown in Ref.~\cite{Alford2017} that
sequential phase transition may lead to the appearance of {\it triplet
configurations} with three stars having the same mass but different radii.
If realized, such a possibility will lead to a proliferation of the possible
combinations of stars that could have been involved in the GW170817 event.
This extends the study of Ref.~\cite{Paschalidis2018} where the tidal
deformabilities of {\it twin configurations} have been analyzed.

This paper is organized as follows. In Sec.~\ref{sec:Construction} we
briefly define the EOS that we use to describe the  hadronic and quark
phases. In Sec.~\ref{sec:Models} we present the results for tidal
deformabilities of spherically symmetric stellar configurations and
in Sec.~\ref{sec:Implications} we confront them with the inferences
from GW170817 and NICER observations of PSR J0030+0451. Our conclusions
are given in Sec.~\ref{sec:Conclusions}.

\section{Construction of EOS}
\label{sec:Construction}

\subsection{Hadronic EOS}

To assess how our results depend on the hadronic matter EOS used to
describe low-density matter, we employ three representative EOS which
feature different hadronic compositions: (i) a purely nucleonic EOS
calculated from the DD-ME2 functional~\cite{Lalazissis2005}, as used
in Ref.~\cite{Alford2017}, which is stiff in the entire relevant density
range; (ii) a hyperonic EOS, based on an extended version of the DD-ME2
functional that includes hyperons, which is softer than the nucleonic
one at densities $\rho/\rho_0 \gtrsim 2.5$~\cite{Lijj2018a}; and (iii)
a hyperon-$\Delta$ admixed matter EOS~\cite{Lijj2018a}, which is soft
at low densities ($1.5 \lesssim \rho/\rho_0 \lesssim 3.0$) but is stiff
at high densities ($\rho/\rho_0 >3.0$).
The tabulated EOS are provided in Table~\ref{tab:EOSH}.

\begin{table}[b]
\caption{
  EOS of hadronic matter used to model hybrid stars. The first
  column lists the density $\rho$ is in units of fm$^{-3}$. The
  columns correspond to $\epsilon-p$ relation for
  purely nucleonic ($N$), hyperonic ($NY$) and hyperon-$\Delta$ admixed (($NY\Delta$)) matter.
  These are given in units of MeV~fm$^{-3}$.
}
\setlength{\tabcolsep}{4.4pt}
\label{tab:EOSH}
\begin{tabular}{ccccccc}
\hline
\hline
\multirow{2}*{$\rho$} & \multicolumn{2}{c}{$N$} & \multicolumn{2}{c}{$NY$} & \multicolumn{2}{c}{$NY\Delta$} \\
\cline{2-7}
      & $\ep$ & $p$ & $\ep$ & $p$ & $\ep$ & $p$ \\
\hline
0.072 &  68.20 &  0.41 &  &  &  & \\
0.100 &  94.94 &  0.80 &  &  &  & \\
0.125 & 118.93 &  1.36 &  &  &  & \\
0.150 & 143.07 &  2.26 &  &  &  & \\
0.175 & 167.40 &  3.76 &  &  &  & \\
0.200 & 192.00 &  6.07 &  &  &  & \\
0.225 & 216.95 &  9.44 &  &  & 216.92 &  8.19 \\
0.250 & 242.33 & 14.04 &  &  & 241.94 &  8.45 \\
0.275 & 268.25 & 20.05 &  &  & 267.04 &  9.97 \\
0.300 & 294.78 & 27.57 &  &  & 292.36 & 13.37 \\
0.325 & 322.00 & 36.66 &  &  & 318.04 & 18.94 \\
0.350 & 349.98 & 47.34 & 349.97 & 46.40 & 344.24 & 26.74 \\
0.375 & 378.78 & 59.63 & 378.57 & 54.73 & 371.08 & 36.71 \\
0.400 & 408.45 & 73.50 & 407.69 & 61.72 & 398.64 & 48.33 \\
0.425 & 439.04 & 88.93 & 437.24 & 68.48 & 426.93 & 59.63 \\
0.450 & 470.57 &105.87 & 467.19 & 75.64 & 455.87 & 70.62 \\
0.475 & 503.09 &124.29 & 497.55 & 83.28 & 485.43 & 82.19 \\
0.500 & 536.62 &144.16 & 528.33 & 91.43 & 515.63 & 94.43 \\
0.550 & 606.78 &188.07 & 591.16 &109.37 & 577.91 &120.92 \\
0.600 & 681.17 &237.36 & 655.72 &129.67 & 642.72 &149.68 \\
0.650 & 759.89 &291.80 & 722.08 &152.49 & 710.02 &179.42 \\
0.700 & 842.99 &351.22 & 790.29 &177.97 & 779.46 &205.85 \\
0.750 & 930.50 &415.47 & 860.42 &206.16 & 850.83 &233.39 \\
0.800 &1022.46 &484.46 & 932.51 &237.01 & 924.03 &260.04 \\
0.850 &1118.87 &558.09 &1006.62 &269.83 & 998.87 &286.30 \\
0.900 &1219.75 &636.29 &1082.57 &299.70 &1075.30 &313.72 \\
0.950 &1325.10 &719.01 &1160.18 &330.14 &1153.28 &342.73 \\
1.000 &1434.92 &806.21 &1239.44 &362.05 &1232.85 &373.58 \\
\hline
\hline
\end{tabular}
\end{table}

\subsection{Quark matter EOS}

To model the EOS of the quark phase(s) we employ a {\it synthetic}
constant-sound-speed (CSS) approach~\cite{Zdunik2013,Alford2013}. We
assume a first-order phase transition manifested as a sharp boundary
between the phases, which is the case when mixed phases are disfavored
by surface tension and electrostatic energy costs~\cite{Alford2001,Palhares2010}.
Here we use the extension of CSS EOS to the case of two sequential phase
transitions by writing~\cite{Alford2017}
\bea\label{eq:EOS}
p(\ep) =\left\{
\begin{array}{ll}
p_{1}, &  \ep_1 < \ep < \ep_1\!+\!\Delta\ep_1 \\[0.5ex]
p_1 + s_1 \bigl[\ep-(\ep_1\!+\!\Delta\ep_1)\bigr],
  & \ep_1\!+\!\Delta\ep_1 < \ep < \ep_2 \\[0.5ex]
p_2, & \ep_2 <\ep < \ep_2\!+\!\Delta\ep_2 \\[0.5ex]
p_2+ s_2\bigl[\ep-(\ep_2\!+\!\Delta\ep_2)\bigr], & \ep > \ep_2\!+\!\Delta\ep_2 \ ,
\end{array}
\right.
\eea
where $p$ is the pressure {and $\ep$ is} the energy density. The
first phase transition from hadronic to quark matter takes place
at $\ep_1, p_1$. The second phase transition within the quark phase,
from Q1 to Q2 (e.g.~from the 2SC to CFL color superconducting phase),
takes place at $\ep_2, p_2$; $s_1$ and $s_2$ are squared sound speeds
in phases 1 and 2. If the energy density at the center of a star is
$\ep_{\rm c} < \ep_2$, there is only a single phase transition in
that star. If $\ep_{\rm c} > \ep_2$ then two phase transitions
take place within the star.
%
\begin{figure}[tb]
\centering
\ifpdf
\includegraphics[width = 0.45\textwidth]{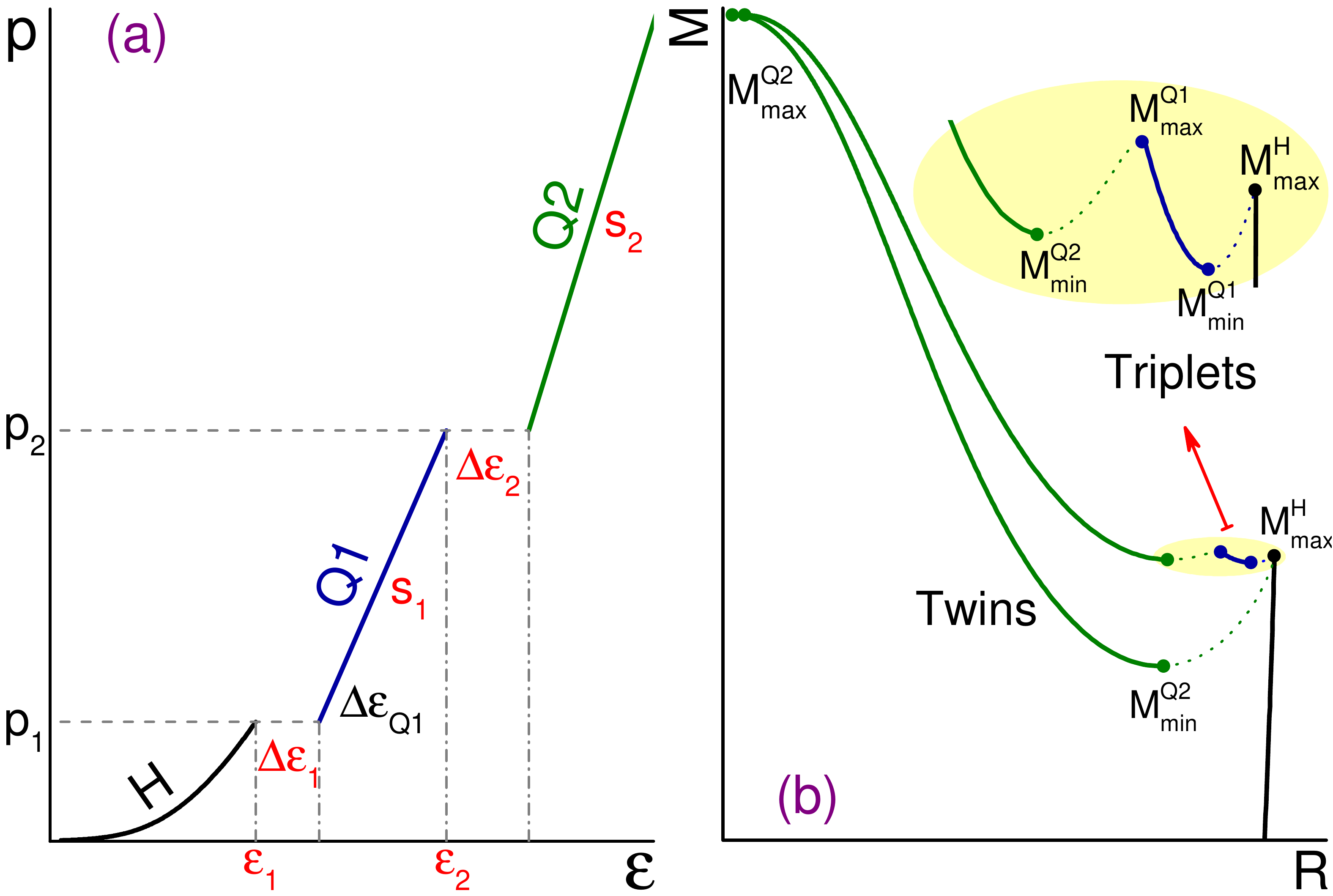}
\else
\includegraphics[width = 0.45\textwidth]{TEM.eps}
\fi
\caption{Schematic plot showing the parametrizations of
the equation of state with two phase transitions (a) and
the resultant mass-radius relation (b).
}
\label{fig:EOS_TEM}
\end{figure}
%

Figure~\ref{fig:EOS_TEM}\,(a) schematically illustrates the
parameterization \eqref{eq:EOS} in the case of double phase
transitions; here H refers to the hadronic phase, Q1, and Q2 to
quark phases. There are, in total, six independent
parameters~\cite{Alford2017}, all of which enter Eq.~\eqref{eq:EOS}
\bea\label{eq:parameters}
\ep_1, \quad \Delta {\ep_1}, \quad
\ep_2, \quad \Delta {\ep_2}, \quad
s_1, \quad s_2\ .
\eea
Instead of $\ep_2$ it is convenient to use the energy-density
width of the Q1 phase, $\Delta\ep_{\text{Q1}}$. A characteristic
mass-radius (MR) relation that results from solving the
Tolman-Oppenheimer-Volkoff equations with the EOS~\eqref{eq:EOS}
as an input is shown in Fig.~\ref{fig:EOS_TEM}\,(b). We illustrate
an EOS with parameters chosen so that twins or triplets of stars
arise, where each new phase of matter introduces a new family of
compact stars. It is convenient for further discussion to define
maximal masses for the branches, labeled by the phase occurring
at the center of the stars in that branch,
\[
M^{\text{H}}_{\text{max}},\quad
M^{\text{Q1}}_{\text{max}},\quad
M^{\text{Q2}}_{\text{max}}\ .
\]
In addition we define the minimum values of the masses of the
stellar branches associated with the quark phases,
\[
M^{\text{Q1}}_{\text{min}},\quad
M^{\text{Q2}}_{\text{min}}\ .
\]
For a large jump in energy density $\Delta\ep_1$, when the central
pressure of a star rises above $p_1$ and Q1 quark matter appears
in the core, the star could become unstable immediately [dashed
line in Fig.~\ref{fig:EOS_TEM}\,(b)]. However, it is possible
to regain stability at higher central pressure (solid line),
resulting in a {\it second stable branch} (or the ``third family''
of compact stars) if the parameters $p_1$, $\Delta\ep_1$ and
$s_1$ are chosen appropriately. In particular, in this case,
twin configurations appear where two stars have the same masses
but different radii~\cite{Paschalidis2018,Alvarez-Castillo2019,
HanSophia2019,Christian2019,HanSophia2019,Montana2019}. If a second
phase transition in the quark phase takes place, then a {\it third
stable branch} (or fourth family) of compact stars containing Q2
quark matter in the core arises. For some choice of parameters
{\it triplet configurations} can arise~\cite{Alford2017} where
three stars have the same masses but different radii.

To study the possible role of quark phases in the merger event
GW170817, we select from the multitude of equilibrium solutions
for the stellar configurations generated by the six-dimensional
parameter space~\eqref{eq:parameters} those that correspond to
low-mass configurations with one or two quark phases. First,
we require the maximum mass, for stars with the densest quark
phase at their center, to reach $2 M_{\odot}$ to be compatible with
Shapiro delay measurements~\cite{Demorest2010,Antoniadis2013,Fonseca2016},
\begin{equation} \label{eq:cond1}
M^{\text{Q2}}_{\text{max}} = 2.0 M_{\odot},
\quad \textrm{(single/double phase transition)}\ ,
\end{equation}
where, in the case of a single phase transition, it is assumed that
the transition takes place directly from the hadronic phase to
the Q2 quark phase. To restrict the parameter space further in
the case of a double phase transition, we impose the following conditions
\bea \label{eq:cond2} M^{\text{H}}_{\text{max}} =
 M^{\text{Q1}}_{\text{max}}, \quad M^{\text{Q1}}_{\text{min}} =
 M^{\text{Q2}}_{\text{min}}, \quad
\textrm{(double phase transition)} \ ,\nonumber\\
\eea
a construction that guarantees that if a twin of hadronic star arises
due to the phase transition to the Q1 phase, the emergence of the new
branch due to the Q2 phase replicates for the triplet the mass range
covered by the twin. Thus, the range of mass twins is automatically
extended to the mass triplets.

Finally, we set $s_1=0.7$ and $s_2=1.0$ as in Ref.~\cite{Alford2017}.
With these constraints we now have a 1-parameter family of stars: if
we specify the value of $M^{\text{H}}_{\text{max}}$ (which is
equivalent to specifying $\ep_1$), all the remaining parameters of the
problem (i.e., $\Delta{\ep_1}$, $\Delta\ep_{\text{Q1}}$ and $\Delta{\ep_2}$)
are determined by the conditions \eqref{eq:cond1} and \eqref{eq:cond2}.
Note that the condition
$M^{\text{Q2}}_{\text{max}}> M^{\text{H}}_{\text{max}}$
is an {\it assumption}; i.e., we require the maximum mass configuration
to occur on the Q2 branch. In the case of a single phase transition,
our discussion is complementary to those given in
Refs.~\cite{Paschalidis2018,Nandi2018,Burgio2018,Christian2019,
Alvarez-Castillo2019,HanSophia2019,Montana2019,Sieniawska2019},
but we take into account the possibility of various compositions
of the hadronic phases surrounding the quark matter core of a hybrid
star, i.e., the hadronic envelope of the star.

\section{Mass, radius and deformability}
\label{sec:Models}

\subsection{Single phase transition}
\begin{figure}[tb]
\centering
\ifpdf
\includegraphics[width = 0.42\textwidth]{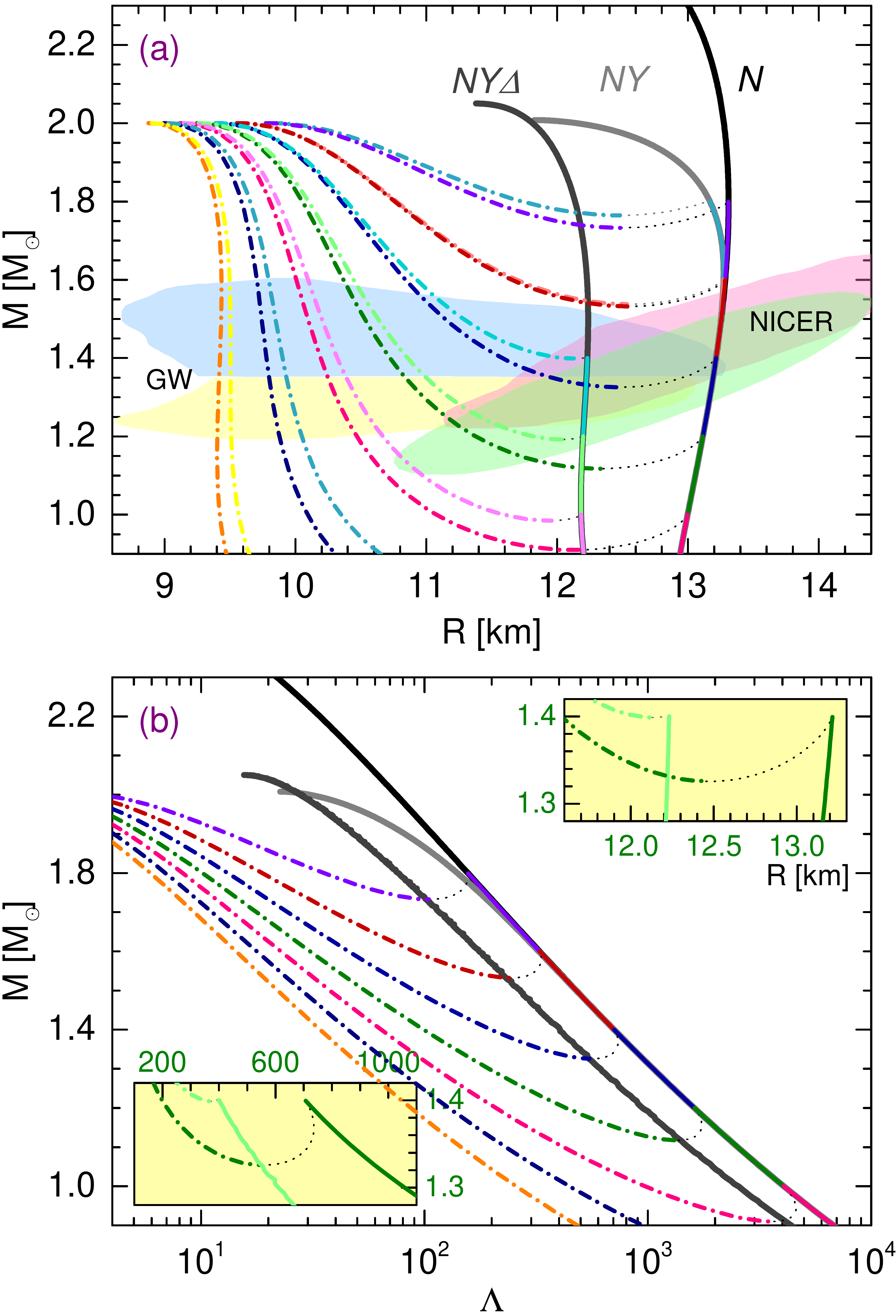}
\else
\includegraphics[width = 0.42\textwidth]{M_RL2.eps}
\fi
\caption{(a) Mass-radius relation for hybrid stars with a single
  phase transition, with three different hadronic envelopes: nucleonic
  $(N)$, hyperonic $(NY)$ and $\Delta$-resonance--hyperon admixed
  $(NY\Delta)$. Each hybrid star branch bifurcates from a hadronic
  sequence at a point which corresponds to the maximum value of
  the mass of the respective hadronic branch $M^{\rm H}_{\text{max}}
  /M_{\odot}$. The figure displays hybrid configurations which
  correspond to $M^{\rm H}_{\text{max}}/M_{\odot} = 0.60$-1.80.
  Note that the hyperons appear only in the two sequences with
  $M _{\text{max}}^{\rm H}/M_{\odot} = 1.60$ and 1.80. In each case
  the maximum mass of the hybrid branch is fixed at $M^{\rm Q2}_{\text{max}}
  /M_{\odot}=2.00$. The dotted lines indicate unstable configurations.
  The shading represents the 90\% posterior credible range of
  masses and radii for the two stars that merged in the GW170817
  event~\citep{LIGO_Virgo2018a}, and the 68\% posterior
  probability of mass-radius for the millisecond pulsar PSR J0030+0451
  obtained by using NICER data~\cite{Riley2019,Miller2019}.
  (b) Mass-deformability relation for
  the configurations shown in (a). We show only the hybrid star
  configurations that have $N$ envelopes for the sake
  of clarity. The insets show the results for the case $M^{\rm
    H}_{\text{max}}/M_{\odot} = 1.40$. The smaller radius
  (deformability) curve corresponds to $NY\Delta$ envelope stars,
  whereas the larger ones correspond to $N$ envelope stars.}
\label{fig:Mass_RL2}
\end{figure}

We start the discussion with the simpler case of hybrid stars with
a single phase transition, varying the composition of the hadronic
envelope. As explained above, our strategy is to vary the value of
$M^{\text{H}}_{\text{max}}$ while keeping $M^{\text{Q2}}_{\text{max}}
/M_{\odot}=2.00$ fixed. The latter condition fully determines the
energy density jump at the first-order phase transition. The
solutions of Tolman-Oppenheimer-Volkoff equations are shown in
Fig.~\ref{fig:Mass_RL2} in the form of MR and mass-deformability
(M$\Lambda$) relations.

\begin{table}[b]
\caption{
 Parameters for the EOS that feature different hadronic envelopes,
 supporting $M^{\rm Q2}_{\text{max}}/M_{\odot}=2.00$. The EOS are
 identified by the maximum mass of the hadronic branch
 $M^{\rm H}_{\text{max}}/M_{\odot} = 0.60$-1.80, in steps of 0.20.
 The energy density at the transition point $\ep_1$ is in units of MeV~fm$^{-3}$.
 The last column lists the mass ranges for twin configurations.
}
\setlength{\tabcolsep}{8.8pt}
\label{tab:EOSa2}
\begin{tabular}{cccccc}
\hline
\hline
Had.  & $M_{\text{max}}^{\rm H}$ & $\rho_{\text{tr}}/\rho_0$ &
$\ep_1$ & $\Delta \ep_1/\ep_1$ & $\Delta M_{\text{twin}}$ \\
\hline
    & 0.600 & 1.555 & 228.431 & 1.813 & 0.095 \\
    & 0.800 & 1.734 & 256.308 & 1.542 & 0.095 \\
    & 1.000 & 1.901 & 283.025 & 1.344 & 0.090 \\
$N$ & 1.200 & 2.069 & 310.423 & 1.187 & 0.082 \\
    & 1.400 & 2.245 & 340.084 & 1.057 & 0.074 \\
    & 1.600 & 2.439 & 373.763 & 0.949 & 0.067 \\
    & 1.800 & 2.665 & 414.552 & 0.859 & 0.065 \\
\hline
\multirow{2}*{$NY$}& 1.600 & 2.549 & 393.077 & 0.857 & 0.062 \\
                   & 1.800 & 3.226 & 516.463 & 0.527 & 0.035 \\
\hline
          & 0.600 & 2.011 & 298.133 & 1.168 & 0.093 \\
          & 0.800 & 2.259 & 337.195 & 0.961 & 0.026 \\
$NY\Delta$& 1.000 & 2.448 & 367.971 & 0.844 & 0.015 \\
          & 1.200 & 2.634 & 399.029 & 0.755 & 0.008 \\
          & 1.400 & 2.881 & 441.724 & 0.653 & 0.001 \\
\hline
\hline

\end{tabular}
\end{table}

It is seen that the onset of hyperons ensures that the maximum masses
of the hadronic sequences stay close to $2M_{\odot}$, whereas the
maximum mass of the pure nucleonic EOS has a much higher value,
$2.48M_{\odot}$. Thus, the hyperonic sequences are consistent with
the {\it upper limit} on the maximum mass inferred from the analysis
of GW170817~\cite{Margalit2017,Shibata2017,Ruiz2018,Rezzolla2018},
whereas their nucleonic counterparts are not. Note, however, that
if a phase transition to quark matter occurs, then the maximum mass
of the purely hadronic branch is not physically relevant. Otherwise,
there are no significant differences between the hyperonic and purely
nucleonic branches of compact stars. The inclusion of $\Delta$-particles
shrinks the radius by about 1\,km as expected~\cite{Lijj2018a,Lijj2019a},
without affecting other features of the MR diagram. The minimum mass
on the hybrid star branch moves to a higher value with the onset of
heavy baryons by less than $10\%$ of mass.

A more quantitative analysis is provided by Table~\ref{tab:EOSa2}
where the transition density, the magnitude of the jump in energy
density at the transition, and the range of masses where twins appear
are shown. This reveals the inverse correlation between the magnitude
of the jump and transition density: the larger the transition density,
the smaller the jump needed to achieve the $2M_{\odot}$ limit for the
hybrid star branch. The range of masses where twin configurations
exist also decreases with the increase of the transition density,
which is the natural consequence of the shape of the MR curves.
We also note that the onset of heavy baryons suppresses the range
of masses over which twin configurations exist.

The mass-deformability relations for the stars with nucleonic matter
envelopes are shown in Fig.~\ref{fig:Mass_RL2}\,(b). The results for
EOS with a hyperonic matter envelope or a hyperon-$\Delta$ admixed matter
envelope are not shown, as they are very close to their nucleonic
counterparts. The key effect of the onset of the quark matter is to
reduce the tidal deformability of the star of a given gravitational
mass, a fact also established in Refs.~\cite{Paschalidis2018,HanSophia2019}.
The reduction is seen in the lower inset of Fig.~\ref{fig:Mass_RL2}\,(b).
This decrease is large enough (several hundreds) that hadronic stars and
hybrid stars will be distinguishable in binary compact star mergers.
Thus, if the masses of the individual stars are measured, their
deformabilities will allow distinguishing between the hadronic
and hybrid stars for a chosen (plausible) hadronic EOS. The tidal
deformability for a 1.4$M_\odot$ star $\Lambda_{1.4}$ can be as low as
$\sim 50$-100, corresponding to a radius of about $\sim 9.5$-10.5~km,
in the extreme case where the phase transitions occur at densities
below 2$\rho_0$. Note, also, that there exists a minimum mass for
neutron star configurations when these are formed in core-collapse
supernovae which is of the order of 1.0$M_{\odot}$~\cite{Schaeffer1983,Ozel2016}.

\subsection{Two sequential phase transitions}
\begin{figure}[tb]
\centering
\ifpdf
\includegraphics[width = 0.42\textwidth]{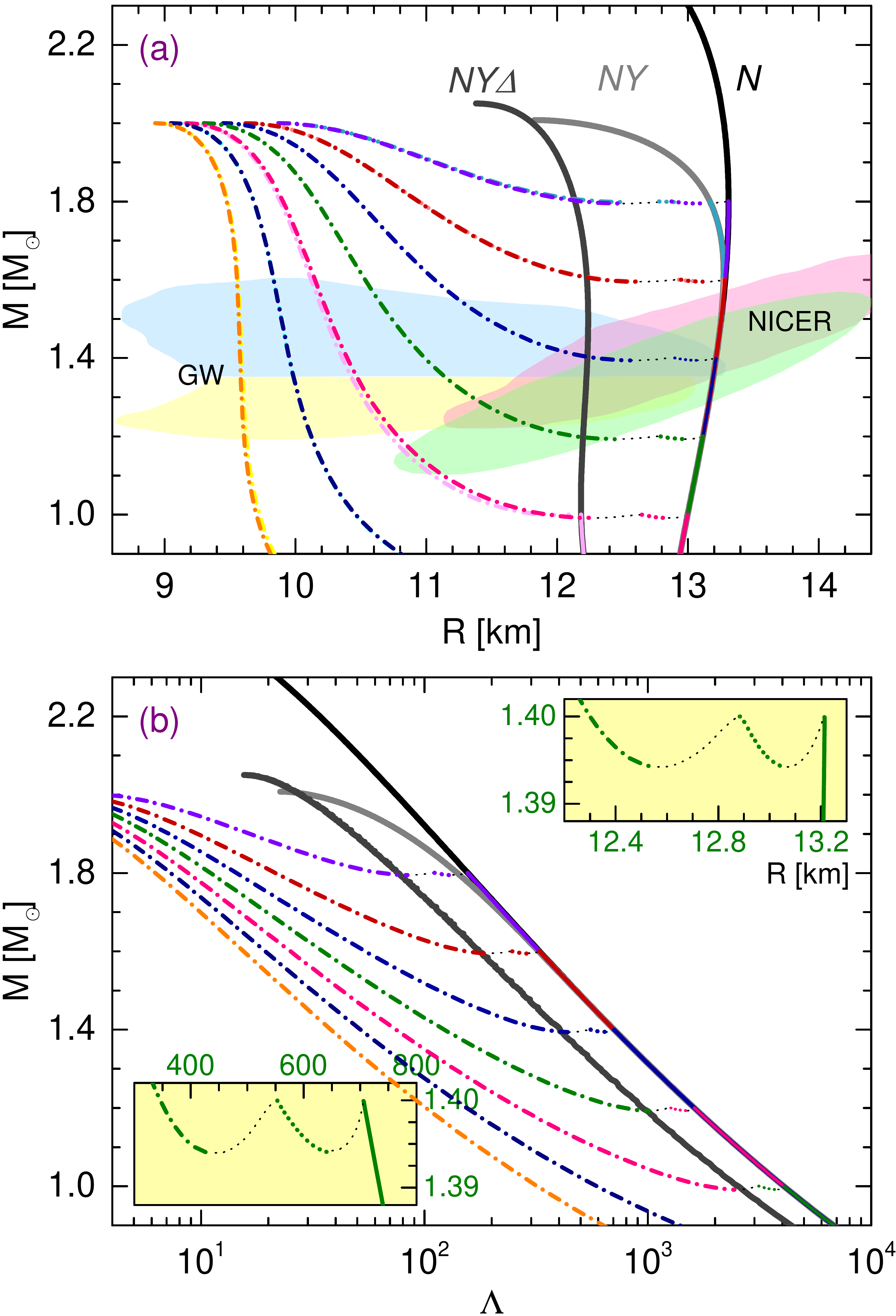}
\else
\includegraphics[width = 0.42\textwidth]{M_RL.eps}
\fi
\caption{ The same as in Fig.~\ref{fig:Mass_RL2}, but in the
  case of quark matter EOS with sequential phase translations.
  The insets show the results for MR and M$\Lambda$ in the
  case of $M^{\text{H}}_{\text{max}}/M_{\odot} = 1.40$. The
  emergence of low-mass triplets in this case is clearly visible.}
\label{fig:Mass_RL}
\end{figure}
%
\begin{table}[b]
  \caption{
  Parameters for the EOS that with different hadronic envelopes,
  supporting $M^{\rm Q2}_{\text{max}}/M_{\odot}=2.00$. The EOS are
  identified by the transition mass $M_{\text{max}}^{\rm H}/M_{\odot}
  = 0.60$-1.80, in steps of 0.20. The energy density at the transition
  point $\ep_1$ is in units of MeV~fm$^{-3}$. The last column presents
  the mass ranges for triplet configurations.
  }
  \setlength{\tabcolsep}{3.0pt}
\label{tab:EOSa}
\begin{tabular}{cccccccccc}
\hline
\hline
Had. & $M_{\text{max}}^{\rm H}$ & $\rho_{\text{tr}}/\rho_0$ &
$\ep_1$ & $\Delta \ep_1/\ep_1$ & $\Delta \ep_2/\ep_1$ &
$\Delta \ep_{\text{Q1}}/\ep_1$ & $\Delta M_{\text{triplet}}$ \\
\hline
\multirow{8}*{$N$}& 0.600 & 1.555 & 228.431 & 0.860 & 0.971 & 0.037 & 0.011 \\
   & 0.800 & 1.734 & 256.308 & 0.808 & 0.755 & 0.042 & 0.010 \\
   & 1.000 & 1.901 & 283.025 & 0.767 & 0.602 & 0.047 & 0.008 \\
   & 1.200 & 2.069 & 310.423 & 0.731 & 0.484 & 0.054 & 0.007 \\
   & 1.365 & 2.213 & 334.670 & 0.706 & 0.405 & 0.062 & 0.006 \\
   & 1.400 & 2.245 & 340.084 & 0.700 & 0.399 & 0.064 & 0.006 \\
   & 1.600 & 2.439 & 373.763 & 0.665 & 0.328 & 0.078 & 0.005 \\
   & 1.800 & 2.665 & 414.552 & 0.630 & 0.292 & 0.109 & 0.005 \\
\hline
\multirow{2}*{$NY$} & 1.600 & 2.549 & 393.077 & 0.604 & 0.294 & 0.075 & 0.004 \\
    & 1.800 & 3.226 & 516.463 & 0.410 & 0.158 & 0.085 & 0.002 \\
\hline
          & 0.600 & 2.011 & 298.133 & 0.658 & 0.522 & 0.026 & 0.004 \\
$NY\Delta$& 0.800 & 2.259 & 337.195 & 0.682 & 0.287 & 0.022 & 0.002 \\
          & 1.000 & 2.448 & 367.971 & 0.679 & 0.173 & 0.021 & 0.001 \\
\hline
\hline
\end{tabular}
\end{table}

Next, we turn to the results found in the case of two sequential phase
transitions. For fixed transitional mass $M^{\text{H}}_{\text{max}}$,
on one hand, triplet configurations occur if the discontinuities in
energy density, $\Delta\ep_1$ and $\Delta\ep_2$, are large enough to
yield {\it separate} {stable} Q1- and Q2-branch solutions. On the
other hand, in order to be compatible with the observational mass
constraint ($M_{\text{max}}/M_{\odot}> 2$)  these jumps have to be
not too large in order for the hybrid branches to be stable. This
restricts the mass range for triplet configurations to a quite narrow
range, as discussed below.

The MR relations for nonrotating stars in this case are shown in
Fig.~\ref{fig:Mass_RL}\,(a). We observe that with
$M^{\text{Q2}}_{\text{max}}/M_{\odot}=2.00$, for the stiff nucleonic
EOS, it is possible to have triplet configurations for a range of
transition densities corresponding to a large range of maximum
masses on the hadronic branch $M^{\rm H}_{\text{max}}/M_{\odot} \in
[0.60-1.80]$. The onset of hyperons softens the high-density domain
of the EOS and affects only the high-mass domain of the MR diagram.
Note that hyperons appear in compact stars with masses
$M/M_{\odot} \geq 1.5$~\cite{Colucci2013,Oertel2015,Lijj2018b}, but
their effect on the mass of the star becomes sizable only for the most
massive stars. The triplet configurations appear in this case for
stars with masses above $1.5M_{\odot}$. The MR curves for stars having
hyperonic or $\Delta$-containing envelopes are almost indistinguishable
from their nucleonic counterparts. Once $\Delta$'s are allowed one finds a
reduction in the radius of the stars, as already observed in the case
of a single phase transition. As a consequence, for $NY\Delta$ stars we
only find triplets at very low mass ($M/M_{\odot} \le 1$) for which the
configurations are unstable. We thus conclude that if the radii on the
hadronic branch are small (in particular, the radius of the maximal-mass
star on that branch, $M^{\rm H}_{\text{max}}$), then the triplet
solutions can exist either only in a narrow range of radii, or they
are eliminated completely. Note that because of the lower mass limit
for neutron stars, the stability of low-mass triplets is an issue.

A more quantitative insight into the results can be obtained from
Table~\ref{tab:EOSa}. It is seen that the triplet solutions exist
in a very narrow range of masses; the presence of heavy baryons
suppresses this range even further. Note also that triplet
configurations arise in the case of nucleonic stars, but are
constrained in the case of hyperon-$\Delta$ admixed matter to
very low values of masses $M/M_{\odot}\le 1$. It may be concluded
from Table~\ref{tab:EOSa} that stiffer hadronic equations of state
can give rise to triplets for a wider range of transition densities.

The mass-deformability relations for the configurations discussed
above are shown in Fig.~\ref{fig:Mass_RL}\,(b) in the case of nucleonic
matter envelopes. The results in the cases of hyperonic envelopes or
hyperon-$\Delta$ admixed envelopes are not shown, as they are very
close to their nucleonic counterparts. However, it should be kept in
mind that in the case of hypernuclear envelopes the triplet
configurations appear in the high-mass range and the deformabilities
are small. In the case of hyperon-$\Delta$ admixed envelopes the
reduction of the radius leads to the restriction to very low masses;
in this case, their tidal deformabilities are large. We again see in
Fig.~\ref{fig:Mass_RL}\,(b) that the phase transition leads to a
reduction of tidal deformability; i.e., the $\Lambda$ value is smaller
for hybrid stars as compared to the same mass nucleonic stars [see
also the lower inset in Fig.~\ref{fig:Mass_RL}\,(b)]. As in the case of
a single phase transition, this decrease is quantitatively large enough,
and the prospects of discrimination between hadronic and hybrid
configurations remain intact. In this case, however, the situation is
more complex as there are pairs of hybrid stars with the same mass.
This leads to a proliferation of the sorts of stars that can take part
in a merger; these issues will be discussed below.

\subsection{Varying $M^{\rm Q2}_{\text{max}}$}
%
\begin{figure}[tb]
\centering
\ifpdf
\includegraphics[width = 0.42\textwidth]{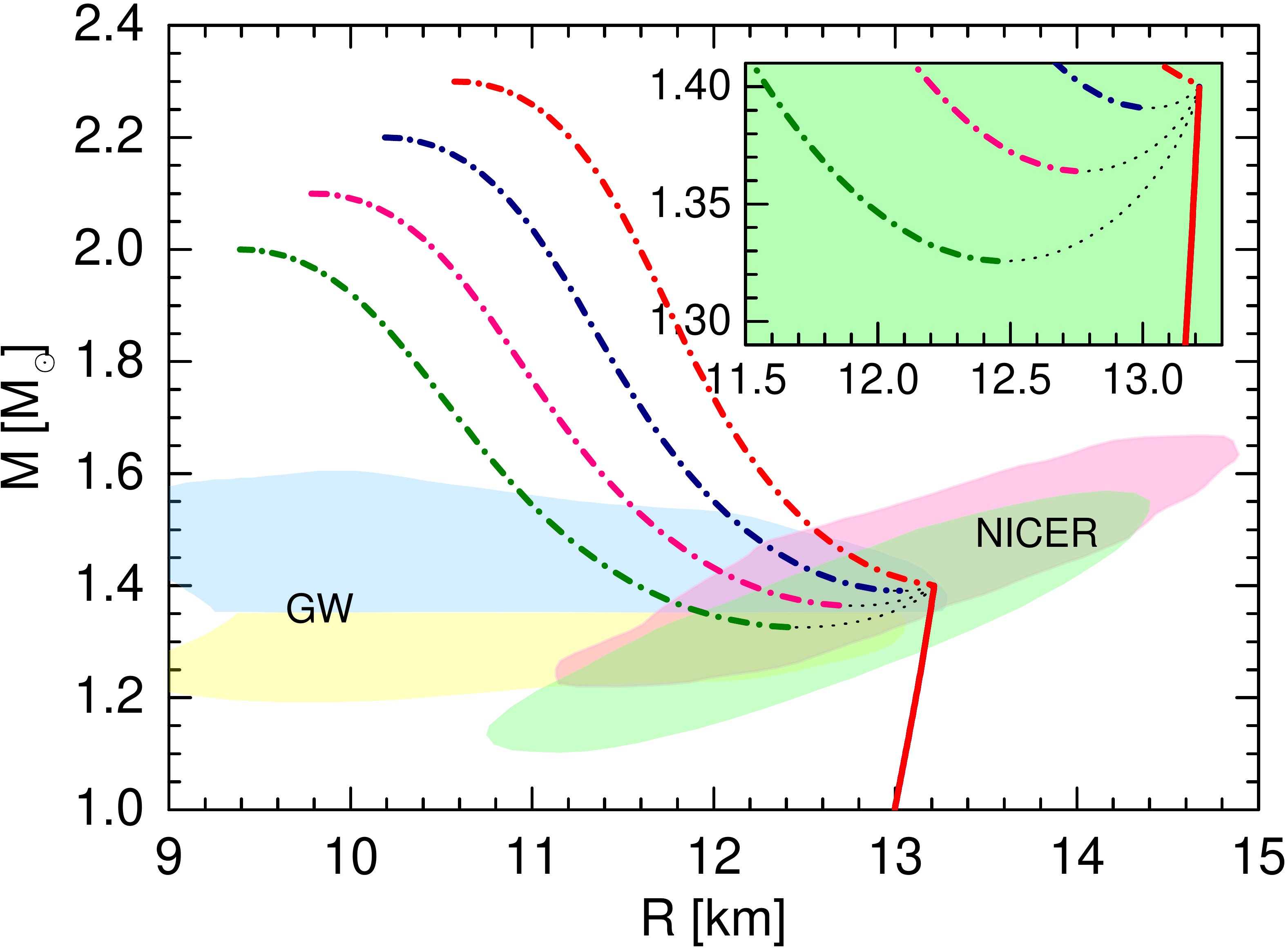}
\else
\includegraphics[width = 0.42\textwidth]{MM_R2.eps}
\fi
\caption{
  Mass-radius relation for hybrid EOS with a single phase transition and
  nucleonic envelope. The EOS are identified by the maximum mass
  $M^{\rm Q2}_{\text{max}}/M_{\odot} = 2.00$-2.30, and the maximum
  mass of the hadronic star, which is fixed at $M^{\rm H}_{\text{max}}
  /M_{\odot}=1.40$. The shading is the same as in Fig.~\ref{fig:Mass_RL2}.
  The emergence of twin configurations is shown in the inset.}
\label{fig:Mass_RM2}
\end{figure}
%
\begin{table}[b]
  \caption{
  Parameters for the EOS with $M^{\rm H}_{\text{max}}=1.40$.
  The EOS are identified by the maximum mass
  $M^{\rm Q2}_{\text{max}}/M_{\odot} = 2.00$-2.30, in steps of 0.10.
  The energy density at transition point $\ep_1$ is 340.084 MeV~fm$^{-3}$.
  The last column presents the mass ranges for twin configurations.
  }
\setlength{\tabcolsep}{19.4pt}
\label{tab:EOSb2}
\begin{tabular}{ccccc}
\hline
\hline
Had. & $M^{\rm Q2}_{\text{max}}$ & $\Delta \ep_1/\ep_1$ &
$\Delta M_{\text{twin}}$ \\
\hline
\multirow{4}*{$N$}& 2.000 & 1.057 & 0.074 \\
   & 2.100 & 0.878 & 0.036 \\
   & 2.200 & 0.721 & 0.009 \\
   & 2.300 & 0.583 & 0.000 \\
\hline
\hline
\end{tabular}
\end{table}
%

So far we have kept the value $M_{\text{max}}^{\rm Q2}/M_{\odot} = 2.00$
fixed. This was in part motivated by the recent arguments which place
the maximum mass of a neutron star in a narrow range slightly above
this value. First, the analysis of the GW170817 event by several
groups suggests an approximate {\it upper limit} on the maximum mass
of a neutron star~\cite{Margalit2017,Shibata2017,Rezzolla2018,Ruiz2018}.
References~\cite{Margalit2017,Shibata2017,Ruiz2018} combined
gravitational-wave and electromagnetic signals with numerical
relativity simulations to limit the maximum mass to the range
2.15-2.30$M_\odot$. The quasiuniversal relations that describe
neutron stars and models of kilonovae were used to draw a similar
bound on the maximum mass in Ref.~\cite{Rezzolla2018}.

%
\begin{figure}[tb]
\centering
\ifpdf
\includegraphics[width = 0.42\textwidth]{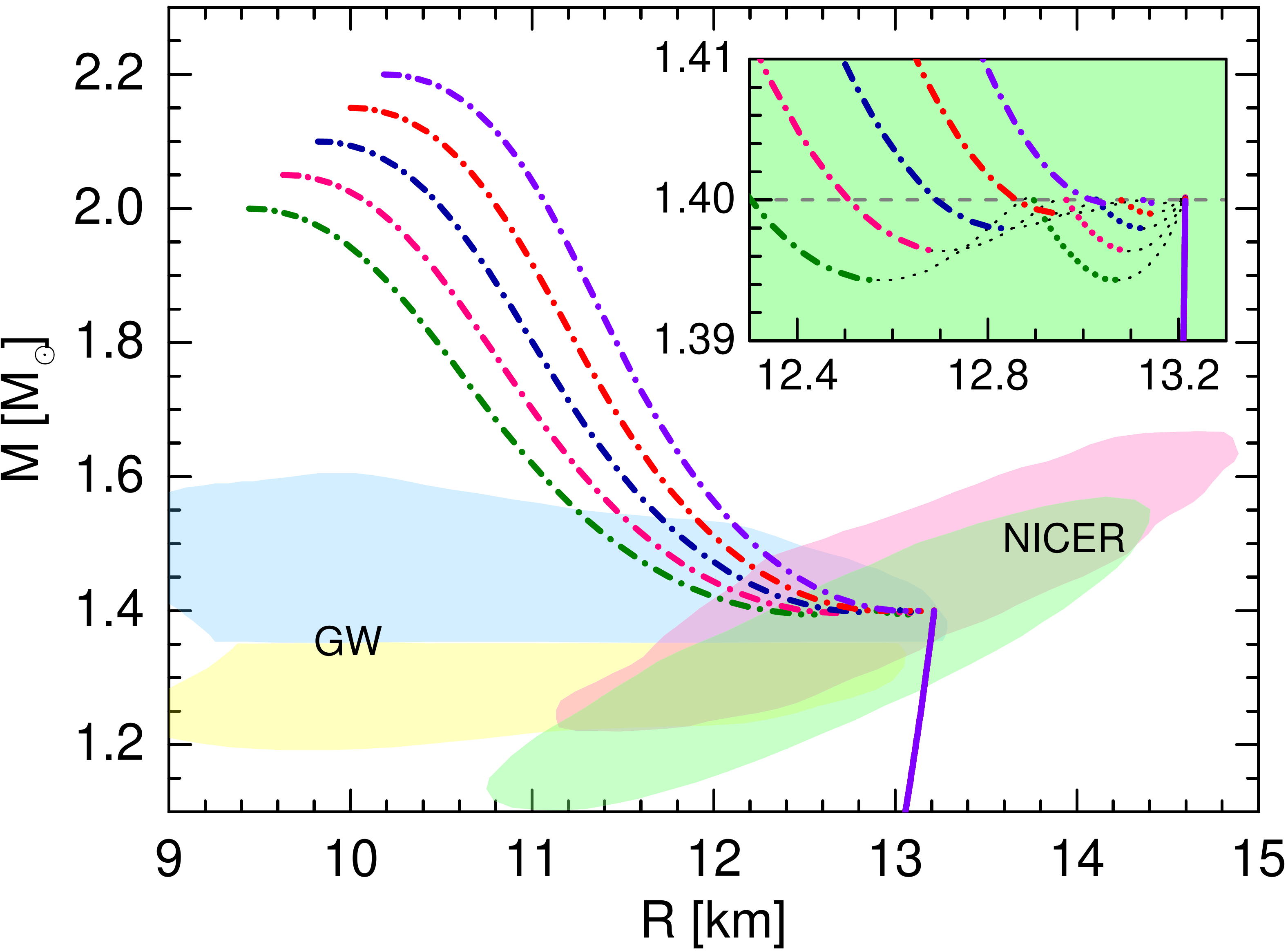}
\else
\includegraphics[width = 0.42\textwidth]{MM_R.eps}
\fi
\caption{ The same as in Fig.~\ref{fig:Mass_RM2}, but calculated
  with hybrid EOS with two sequential phase translation. The inset
  illustrates the emergence of triplet configurations. }
\label{fig:Mass_RM}
\end{figure}
%
\begin{table}[t]
  \caption{
  Parameters for the EOS with $M^{\rm H}_{\text{max}} =1.40$.
  The EOS are identified by the maximum mass
  $M^{\rm Q2}_{\text{max}}/M_{\odot} = 2.00$-2.20, in steps of 0.05.
  The energy density at transition point $\ep_1$ is 340.084 MeV~fm$^{-3}$.
  The last column presents the mass ranges for triplet configurations.
  }
\setlength{\tabcolsep}{8.2pt}
\label{tab:EOSb}
\begin{tabular}{ccccccc}
\hline
\hline
Had. & $M^{\rm Q2}_{\text{max}}$ & $\Delta \ep_1/\ep_1$ &
$\Delta \ep_2/\ep_1$ & $\Delta \ep_{\text{Q1}}/\ep_1$
& $\Delta M_{\text{triplet}}$ \\
\hline
   & 2.000 & 0.700 & 0.392 & 0.064 & 0.006 \\
   & 2.050 & 0.681 & 0.309 & 0.053 & 0.004 \\
$N$& 2.100 & 0.665 & 0.232 & 0.042 & 0.002 \\
   & 2.150 & 0.652 & 0.158 & 0.033 & 0.001 \\
   & 2.200 & 0.642 & 0.088 & 0.025 & 0.000 \\
\hline
\hline
\end{tabular}
\end{table}
%

As mentioned in the Introduction, a direct astrophysical {\it lower bound}
$2.14^{+0.10}_{-0.09}M_\odot$ (68.3\% credibility interval) on the maximum
mass of a neutron star was recently obtained via combination of the Shapiro
delay data taken over 12.5\,yr at the NANOGrav with orbital-phase-specific
observations using the Green Bank Telescope from the mass measurement of
the millisecond pulsar PSR J0740+6620~\cite{Cromartie2019}. Combining the
lower and upper bounds quoted above, one concludes that the maximum mass
of a neutron star is located in the band $2.15$-2.30~$M_{\odot}$.

We next explore the sensitivity of our results towards varying the
value of $M_{\text{max}}^{\rm Q2}$ in the cases of single and two
sequential phase transitions. To this end, we have fixed the value of
$M^{\rm H}_{\text{max}}/M_{\odot}=1.40$ and imposed again the
conditions~\eqref{eq:cond2} adapted for the cases of double phase transition.

The effects of varying $M_{\text{max}}^{\rm Q2}$ on the MR relations are
shown in Fig.~\ref{fig:Mass_RM2} for a single phase transition and in
Fig.~\ref{fig:Mass_RM} for two sequential phase transitions. We observe
that a larger maximum mass leads to smaller ranges of radii (and masses)
where twin and triplet solutions exist. This effect is illustrated
more quantitatively in Tables~\ref{tab:EOSb2} and~\ref{tab:EOSb} where
additional parameters fully characterizing the EOS are given.

\section{Comparison with the data from GW170817 event}
\label{sec:Implications}

\begin{figure}[htb]
\centering
\ifpdf
\includegraphics[width = 0.45\textwidth]{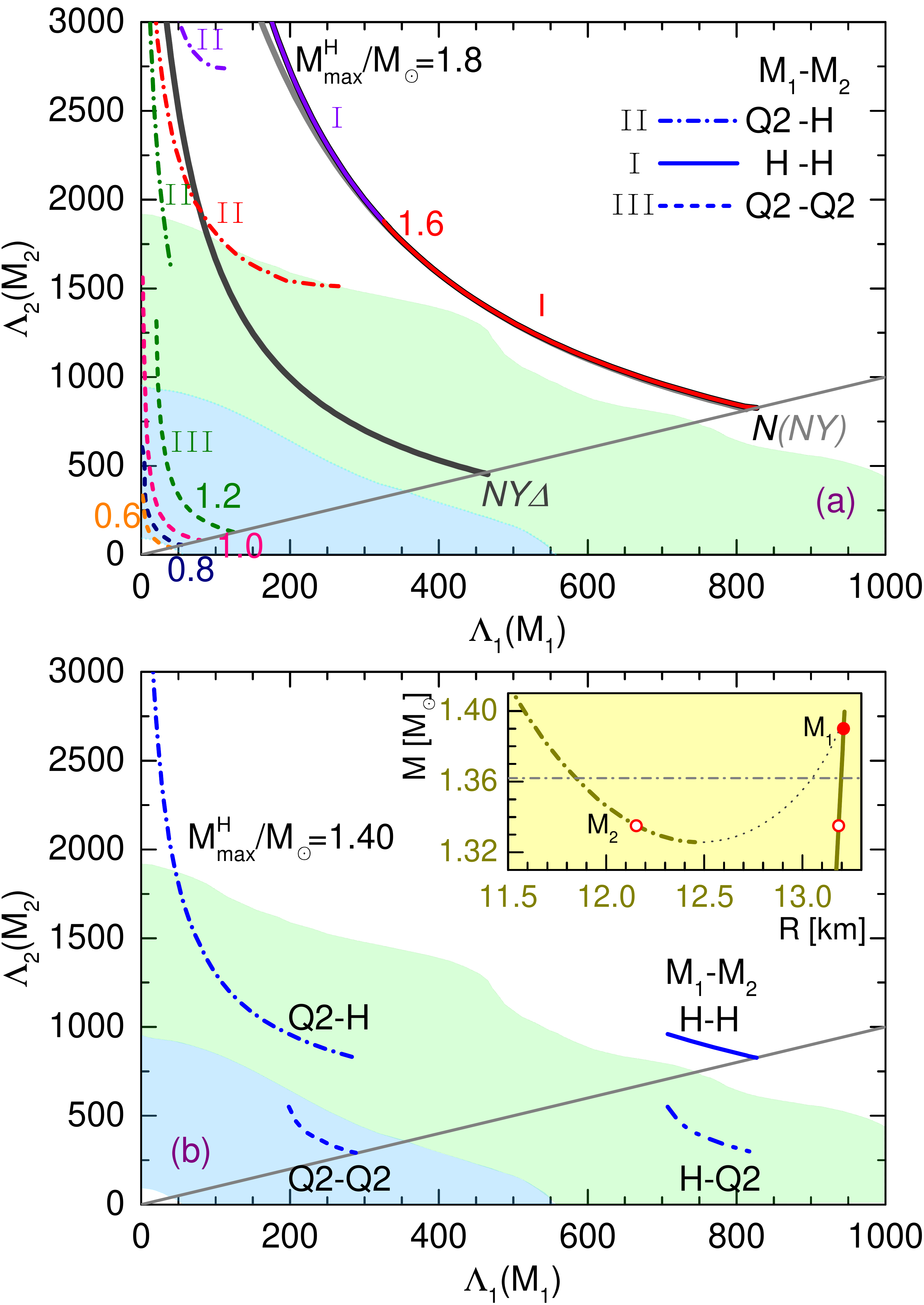}
\else
\includegraphics[width = 0.45\textwidth]{GW_LL2.eps}
\fi
\caption{
  (a) Tidal deformabilities of compact objects with a single
  phase transition for a fixed value of binary chirp mass
  $\mathcal{M} = 1.186M_\odot$. The EOS are labeled by the
  maximal mass $M^{\rm H}_{\text{max}}$ of the hadronic branch
  of the sequences. The three types of pairs for stars with
  masses $M_1$ and $M_2$ are H-H (solid lines), Q2-H
  (dash-dotted lines), and Q2-Q2 (dashed lines). These three
  possibilities are also labeled by roman numerals I, II and III.
  (b) The same as in (a) but for fixed$M^{\rm H}_{\text{max}}
  = 1.40M_\odot$, which is close to the value $M_1 = M_2 =
  1.362M_{\odot}$ inferred from the GW170817 event assuming it
  involved equal-mass stars. The shaded regions correspond
  to the 50\% and 90\% credibility regions taken from the
  analysis of GW170817 within the PhenomPNRT model~\citep{LIGO_Virgo2018b}.
  The inset shows the mass-radius relation around the phase
  transition region. The open circles (labeled $M_2$) are the
  masses of two possible companions for the star of mass
  $M_1$ (full circle) for a fixed value of binary chirp mass
  $\mathcal{M} = 1.186M_\odot$.}
\label{fig:Lam_Lam2}
\end{figure}

\begin{figure}[htb]
\centering
\ifpdf
\includegraphics[width = 0.45\textwidth]{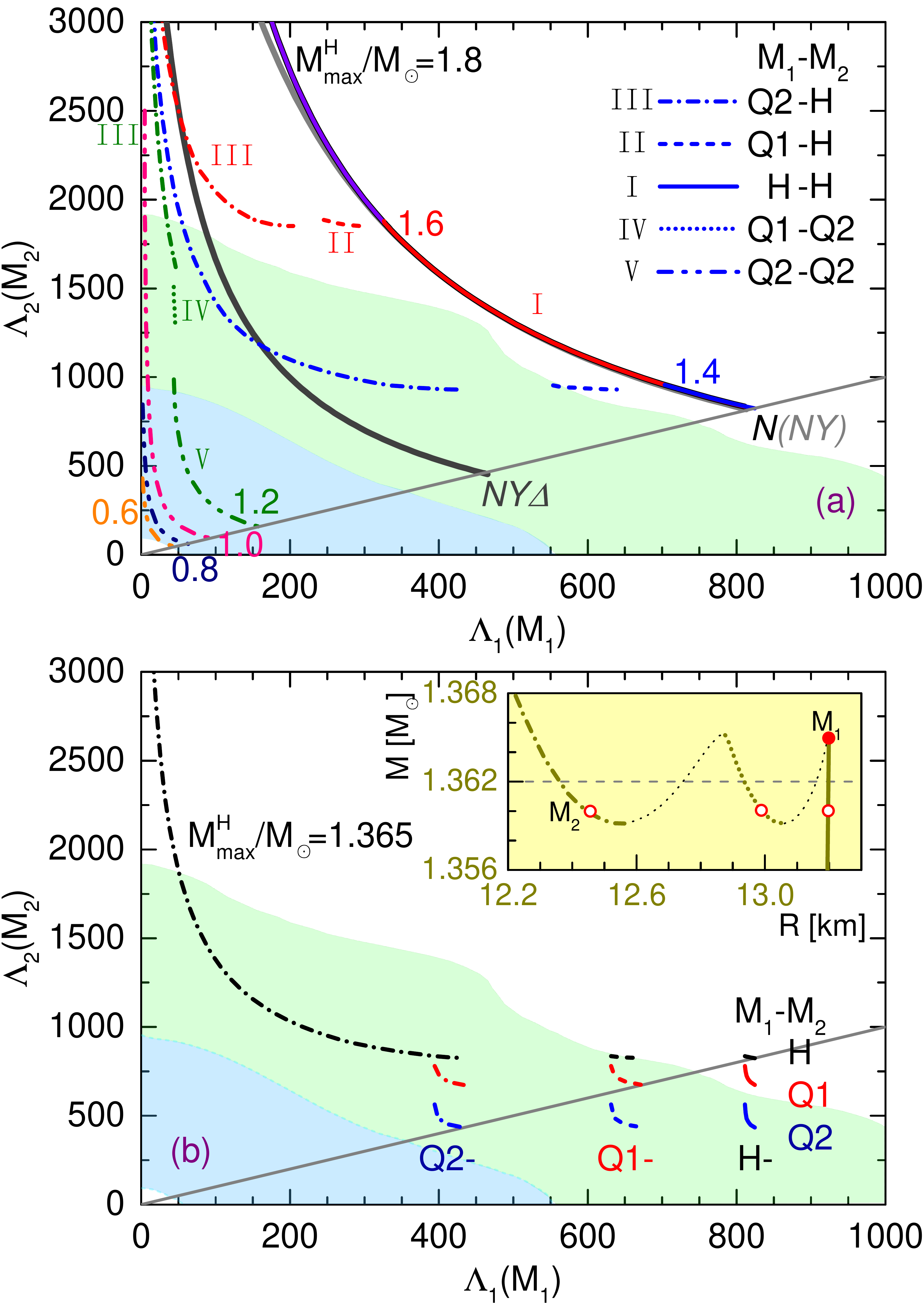}
\else
\includegraphics[width = 0.45\textwidth]{GW_LL.eps}
\fi
\caption{ The same as in Fig.~\ref{fig:Lam_Lam2}, but in the case
  of two sequential phase translation (the label Q2 in the case
  indicates stars with two phase transitions, whereas Q1 - with
  a single phase transition). The distinct types of pairs of
  neutron stars are labeled as H-H (solid line), Q1-H (dashed line),
  Q2-H (dash-dotted line), Q1-Q2 (dotted line), and Q2-Q2
  (dash-double-dotted line). These are also labeled by Roman numerals
  from I to V in the given order. Panel (b) in this case contains the
  following cases: (i) the primary is a Q2 star (dashed-dotted lines)
  with the secondary being H (black), Q1 (red), and Q2 (blue) stars; (ii)
  the primary is a Q1 star (dashed lines) with secondary being H
  (black), Q1 (red), and Q2 (blue) stars; and, finally, the primary is
  a hadronic star (solid lines) with the secondary being H (black),
  Q1 (red) and Q2 (blue) star. The inset in panel (b) shows
  the mass-radius relation, where the open circles correspond to masses
  (labeled $M_2$) of the three possible companions of the primary
  star with mass $M_1$ (full circle) for a fixed value of binary
  chirp mass $\mathcal{M} = 1.186M_\odot$. }
\label{fig:Lam_Lam}
\end{figure}

We now confront the tidal deformabilities of our neutron star models
with the observational constraints for this quantity obtained from
the analysis of the GW170817 event. We assume the chirp mass
$\mathcal{M} = 1.186 M_\odot$ and compare only with the analysis which
assumes the (more plausible) low-spin case~\cite{LIGO_Virgo2018a}.
Figure~\ref{fig:Lam_Lam2} displays the tidal deformabilities
$\Lambda_1$ and $\Lambda_2$ of the stars involved in the binary with
masses $M_1$ and $M_2$ in the case of a single phase transition.
In panel (a) the hybrid EOS for the binary are identified by the
value of $M_{\rm max}^{\rm H}$. In panel (b) we fix this value
at $1.40M_{\odot}$, which is close to the experimentally inferred
value in the case of equal-mass binary assumption. The same
in the case of two sequential phase transitions is shown in
Fig.~\ref{fig:Lam_Lam}. In these figures, the diagonal line
corresponds to the case of an equal-mass binary with
$M_{1,2} = 1.362M_\odot$. The shaded areas correspond to the
90\% and 50\% confidence limits, which are inferred from the
analysis of the GW170817 event using the PhenomPNRT waveform
model~\citep{LIGO_Virgo2018b}. Note that the observational analysis
depends on the assumed theoretical waveforms. For example, 90\%
confidence upper limits on tidal deformabilities predicted by the
TaylorF2 model are larger by about 20\% than those predicted by
the PhenomRNRT model used in our figures.

It is seen that the largest values of the deformability in the
$\Lambda_1$-$\Lambda_2$ plane are generated by the curve corresponding
to the purely nucleonic EOS, as this is the hardest possible EOS in
our collection of models. The case of hyperonic EOS is almost
indistinguishable from the nucleonic one. The corresponding curves
remain slightly outside the 90\% credibility level set by the PhenomPNRT
model ($\sim10\%$ for the equal-mass case). In the case where
$\Delta$'s are included in the composition, the $\Lambda$ values
are significantly reduced in agreement with the previous
result~\cite{Lijj2019a}.

%
%
%

The fact that mass twins are allowed in the case of the single phase
transition implies that, in general, two types of pairs of neutron stars
could be involved in a merger event, namely, H-H, Q2-H for models with
$M_{\rm max}^{\rm H} > 1.45M_{\odot}$, and Q2-H, Q2-Q2 for EOS with
$M_{\rm max}^{\rm H} < 1.35M_{\odot}$, where H and Q2 denote hadronic
and hybrid stars (with the Q2 quark phase); see Fig.~\ref{fig:Lam_Lam2}\,(a).
As a general trend, the H-H combination produces the largest tidal
deformabilities, the Q2-H combination produces intermediate values of
tidal deformabilities, and the smallest values are obtained for the
Q2-Q2 combination. Since low values of tidal deformabilities are
favored by the observational analysis, one may conclude that the phase
transition to quark matter helps to reduce the tension between
purely hadronic models and the data (see also
Ref.~\cite{Paschalidis2018}). Figure~\ref{fig:Lam_Lam2} also
demonstrates the dependence of the results on the transition density
$\ep_1$ parametrized in terms of $M_{\rm max}^{\rm H}$: for
lower phase transition density (i.e., a smaller $M_{\rm max}^{\rm H}$
value and larger quark core) the tidal deformabilities are smaller, in
agreement with the fact that the stars are more compact. Furthermore,
purely hadronic members can be obtained only if the transition density
is high enough so that twin stars have masses heavier than
$1.362M_{\odot}$. Interestingly, in the case where
$M_{\rm max}^{\rm H} = 1.40M_{\odot}$, the mass twins appear around
$1.362M_{\odot}$, as shown in the inset of Fig.~\ref{fig:Lam_Lam2}\,(b).
Furthermore, an additional H-Q2 curve lies below the diagonal,
which implies that for $M_1 \geqslant M_2$ one has $\Lambda_1\geqslant\Lambda_2$,
which is contrary to the expectation that $\Lambda_1\leqslant\Lambda_2$,
as already observed by several authors~\cite{Christian2019,
Alvarez-Castillo2019,HanSophia2019,Montana2019,Sieniawska2019}.

If mass triplets are allowed in the case of two sequential phase
transitions, there are, in general, three types of pairs of neutron stars
that could be involved in a merger event, namely, H-H, Q1-H, and Q2-H
for EOS with $M_{\rm max}^{\rm H} > 1.37M_{\odot}$, and Q2-H,
Q1-Q2 and Q2-Q2 for EOS with $M_{\rm max}^{\rm H} < 1.36M_{\odot}$;
see Fig.~\ref{fig:Lam_Lam}\,(a). For $M_{\rm max}^{\rm H} = 1.365M_{\odot}$
the mass triplets arise around $1.362M_{\odot}$. In this case,
the number of possible pairs proliferates to a total of $3!=6$.
Three of these are pairs of stars of the same type --- H-H, Q1-Q1,
and Q2-Q2 --- and three are pairs of mixed type --- H-Q1, H-Q2, Q1-Q2;
see Fig.~\ref{fig:Lam_Lam}\,(b). This highlights the fact that in
the case of two first-order sequential phase transition it is possible
to generate a rich variety of compact star mergers. In this case,
we again note that three curves, namely H-Q1, H-Q2, and Q1-Q2,
lie below the diagonal line in Fig.~\ref{fig:Lam_Lam}\,(b).
This is a direct extension of the effect already seen in the
case of twins to the case of triplet stars.

%
\begin{figure}[b]
\centering
\ifpdf
\includegraphics[width = 0.42\textwidth]{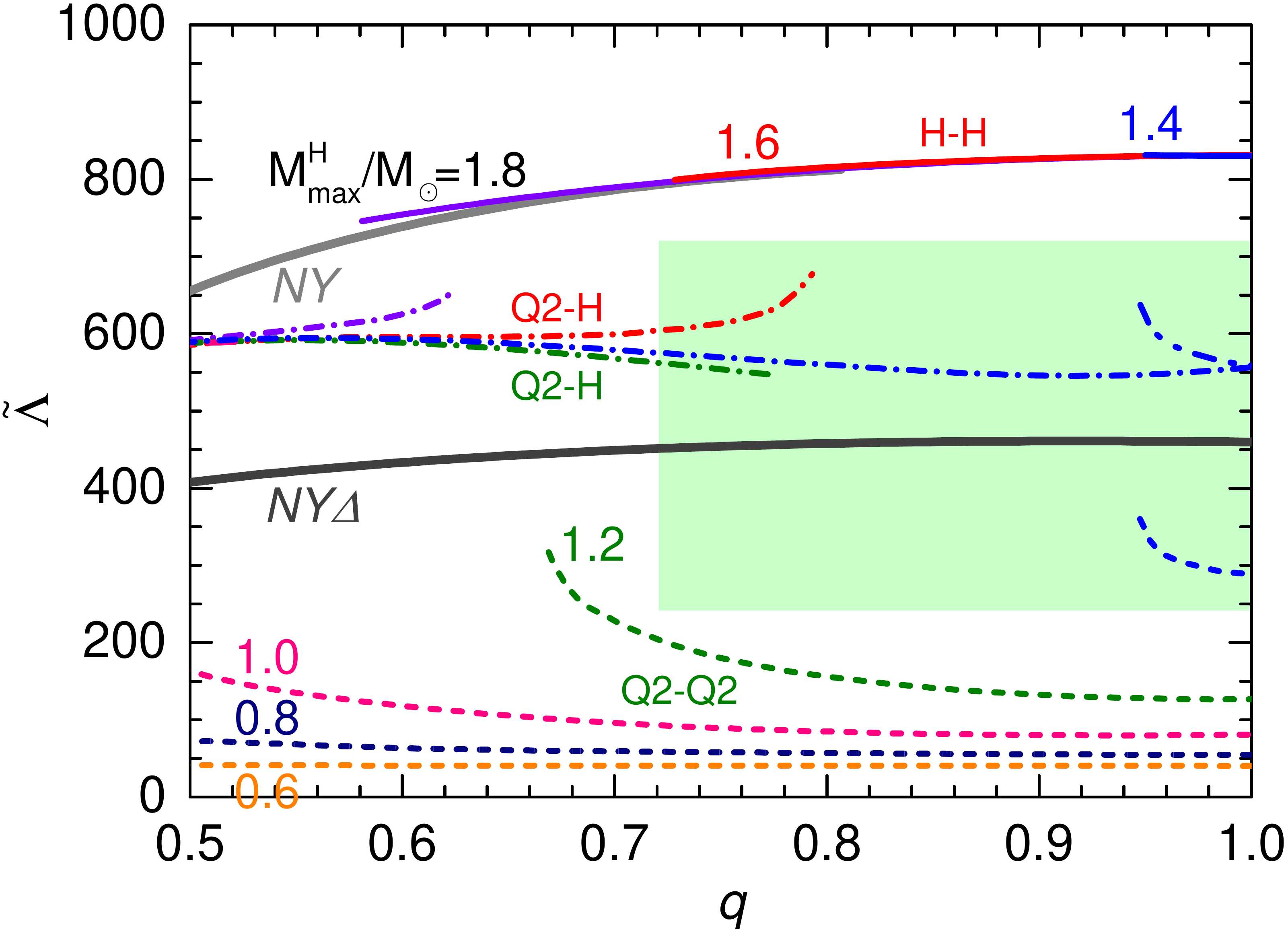}
\else
\includegraphics[width = 0.42\textwidth]{WL_Q2.eps}
\fi
\caption{ Mass weighted deformability vs. mass asymmetry for a binary
  system with fixed chirp mass $\mathcal{M} = 1.186 M_\odot$ predicted
  by a range of hybrid EOS with a single phase transition and various
  values of $M_{\text{max}}^{\rm H}$. The labeling is the same as
  in Fig.~\ref{fig:Lam_Lam2}. The error shading indicates the constraints
  estimated from the GW170817 event~\cite{LIGO_Virgo2018b} and the
  electromagnetic transient AT2017gfo~\cite{Coughlin2018,Radice2019,Kiuchi2019}. }
\label{fig:WLam_Q2}
\end{figure}
%
\begin{figure}[tb]
\centering
\ifpdf
\includegraphics[width = 0.42\textwidth]{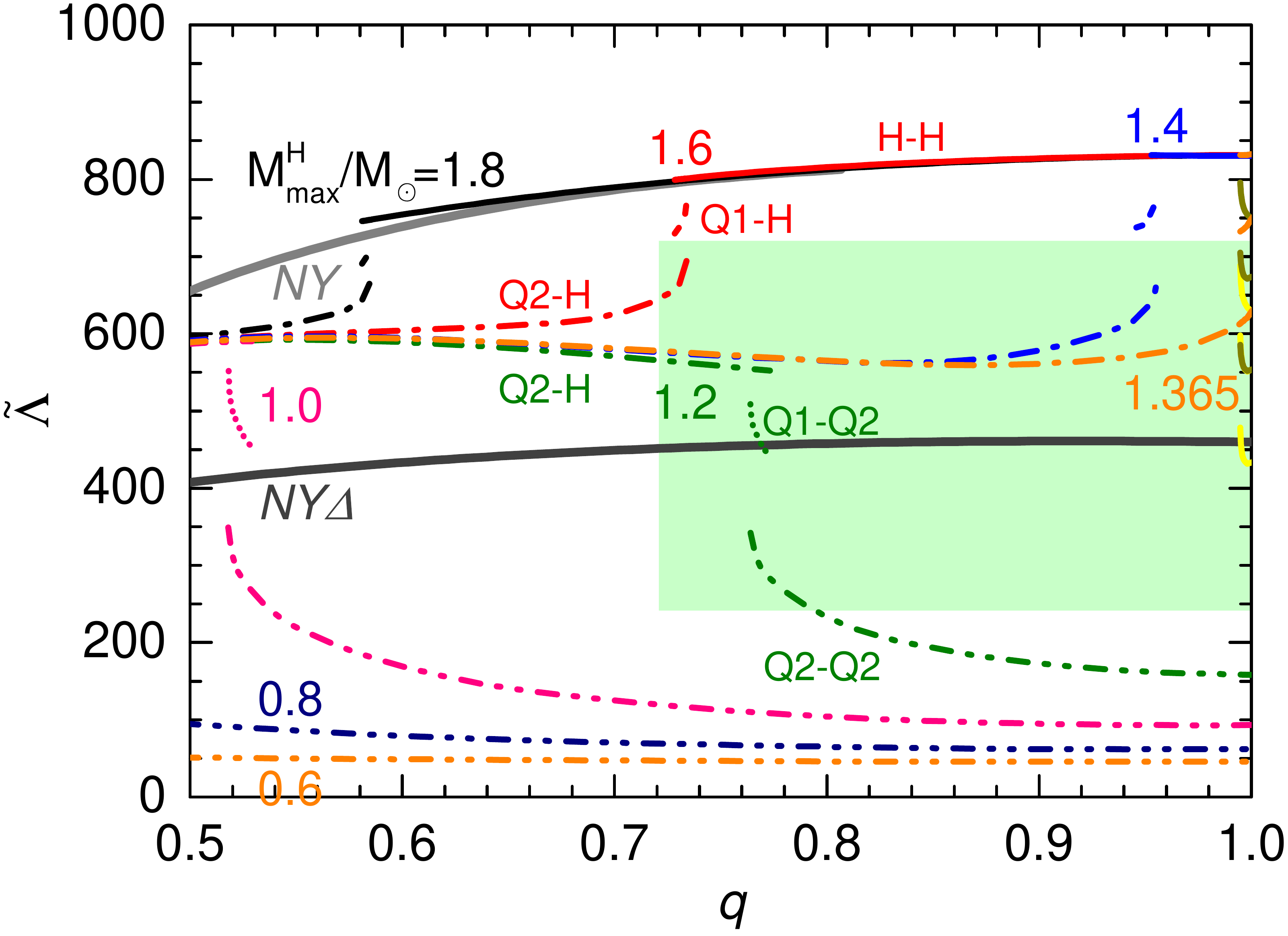}
\else
\includegraphics[width = 0.42\textwidth]{WL_Q.eps}
\fi
\caption{ The same as in Fig.~\ref{fig:WLam_Q2}, but in the case
  of EOS with two sequential phase translations. The labeling is
  the same as in Fig.~\ref{fig:Lam_Lam}. }
\label{fig:WLam_Q}
\end{figure}

A quantity that has been accurately measured for the GW170817 event during
the inspiral phase is the mass weighted average tidal deformability
$\tilde{\Lambda}$~\cite{Hinderer2010,Wade2014}, (for the measurements
result, see Refs~\cite{LIGO_Virgo2017c,LIGO_Virgo2018a,LIGO_Virgo2018b}).
We show this quantity in Figs.~\ref{fig:WLam_Q2} and~\ref{fig:WLam_Q}
as a function of the ratio of masses $q = M_2/M_1$ of merger components
for a fixed chirp mass $\mathcal{M} = 1.186 M_\odot$. The ratio $q$
in the case of the GW170817 event was inferred to be in the range
$0.72\le q\le 1$~\cite{LIGO_Virgo2018b,Coughlin2018}. The boundaries
on $\tilde{\Lambda}$, which were set by the gravitational-wave and
electromagnetic observations, respectively, are also shown. The upper
limit $\tilde{\Lambda} = 720$ was set by the analysis of the GW170817
event (more precisely, this is the lowest value for this limit at the 90\%
confidence level~\cite{LIGO_Virgo2018b}). The lower limit
$\tilde{\Lambda} \approx 250$ was deduced from the analysis of the
electromagnetic counterpart of GW170817, e.g.,
AT2017gfo~\cite{Radice2018,Coughlin2018,Radice2019,Kiuchi2019}.
This lower limit on $\tilde{\Lambda}$ indicates approximately that
the radius for a canonical neutron star must be larger than
$\sim10.5$~km~\cite{Radice2019, Kiuchi2019}.

From Figs.~\ref{fig:WLam_Q2} and~\ref{fig:WLam_Q} we see that
$\tilde{\Lambda}$ weakly depends on the mass ratio $q$ for purely
hadronic EOS as well as for hybrid EOS with $M^{\rm H}_{\text{max}}
/M_{\odot} < 1.0$. For hybrid EOS with larger $M^{\rm H}_{\text{max}}$,
some variation is observed in the range of several hundreds.
It is further seen that the configurations with phase transition(s),
which are more compact than their nucleonic counterparts provide
$\tilde \Lambda$ values which are favored by the limits deduced
from the GW170817 event alone. The hybrid EOS with transitional masses
in the range $1.2\lesssim M_{\text{max}}^{\rm H}/M_{\odot}\lesssim1.6$
are roughly consistent with above-mentioned limits. On the other hand,
the low-mass models with $M_{\text{max}}^{\rm H}/M_{\odot} \lesssim 1.0$
can be excluded on the basis of the electromagnetic observations.

It is expected that more neutron star binary mergers will be observed
within the coming years. Now, imagine comparing two such events in which
the observed values of $\Lambda_1$ are close but the $\Lambda_2$ values
are different. Such a situation may arise when the secondary star $M_2$
has a twin or triplet configuration. Such an observation would be evidence
for the existence of a strong first-order phase transition at low transition
density (or mass). If, conversely, the same value for $\Lambda_2$ but
different values for $\Lambda_1$ are observed, (i.e., two stars belonging to twin
or triplet combinations with the primary star's mass $M_1$ were merged in each
event), then this would be an indication for a high transition density;
see Figs.~\ref{fig:Lam_Lam2} and~\ref{fig:Lam_Lam}. Furthermore, for two events
in which the observed values of chirp mass $\mathcal{M}$ are similar but
the values of mass-weighted deformabilities $\tilde \Lambda$ are significantly
different, such an observation also serves as a signal for the existence of
(sequential) strong first-order phase transition(s); see Figs.~\ref{fig:WLam_Q2}
and~\ref{fig:WLam_Q}.

\section{Conclusions}
\label{sec:Conclusions}

In this work we have explored several compositions of compact
stars. On the hadronic side, we considered purely nucleonic,
hypernuclear and $\Delta$-resonance plus hypernuclear matter.
The hadronic matter was described within the density functional
theory. We further assumed a first order phase transition to
quark matter characterized either by a single phase or two phases.
All phase transitions were assumed to be first-order and involved
a density jump at the phase transition. The quark matter was
described using the constant speed of sound parametrization, each
phase having its own speed of sound. Starting from the EOS of these
phases, we constructed the stellar models, and obtained their masses,
radii and tidal deformabilities. Our results can be summarized
as follows:

\smallskip\noindent\underline{(a) Variations in the hadronic envelope of the models:}\\
We observe, consistent with the previous work, that the onset of
hyperons allows us to keep the maximum mass of our configurations
below the upper limit on the maximum mass, $\sim 2.2 M_{\odot}$, of
compact star sequences deduced from the GW170817 event. As shown
previously, the inclusion of $\Delta$'s reduces the radii of the
configuration with intermediate masses and reduces considerably
the tidal deformabilities. Allowing for phase transition(s), we
recover the general features known for hybrid stars, in particular,
the occurrence of twin and triplet configurations. Interestingly,
the low radii obtained in the case of $\Delta$-admixed matter exclude
the appearance of stable stars with two phases of quark matter for
a wide density range and, consequently, the possibility of triplets.

\smallskip\noindent\underline{(b) Conditions for the occurrence of low-mass twins/triplets:}\\
Previous work has shown that twins are possible for masses
$M \sim 1.4M_{\odot}$~\cite{Paschalidis2018}, while triplets were found for
$M\sim 1.7M_{\odot}$~\cite{Alford2017}. Here we show that if the
H-Q1 transition density is taken to be sufficiently low, then
low-mass triplet stars can be obtained for nucleonic and
hyperon-$\Delta$ admixed envelopes. For models, we produce families
of stars whose maximum mass (attained on the Q2 branch) is $2M_\odot$.
In doing so we assumed that the quark matter phases are characterized
by stiff EOS with speeds of sound $s_1= 0.7$ (Q1 phase) and $s_2 = 1$
(Q2 phase) as has been assumed in Ref.~\cite{Alford2017}.

\smallskip\noindent\underline{(c) Tidal deformabilities and interpretation of GW170817:}\\
Perhaps the most striking consequence of the existence of triplets
is that the number of possible pairs that could be involved in a binary
merger proliferates to $6$; explicitly the pairs can now include H-H,
an H-hybrid (Q1 or Q2), and a hybrid (Q2 or Q1)-hybrid (Q1 or Q2)
binaries. For equal-mass binaries, all six possibilities can occur
in nature, although the range of masses where this can occur is narrow.
A generic conclusion, independent of the specific components of the
binary, is that the emergence of quark phase(s) lifts the tension
between the observations and the tidal deformabilities obtained
for purely hadronic stars constructed based on relativistic
density functional theory.

\smallskip\noindent\underline{(d) Inferring first-order phase transitions from observations:}\\
The first analysis of NICER data provided values of the radius
of a canonical $M\sim 1.4M_{\odot}$ star with unprecedented accuracy
~\cite{Riley2019,Miller2019}. At the current level of accuracy
these results do not contradict the radii inferred from the GW170817
event, as the values of radii extracted from both experiments have
an overlap region, see Fig.~\ref{fig:Mass_RL2}. Should future
analysis of these data or future observations result in
non-overlapping regions for the radii of compact objects, this would
indicate the existence of two classes of neutron stars with the same
masses but substantially different radii - a hallmark of twins or
triplet configurations.

Given the possibility that the LIGO-Virgo Collaboration will observe a few
neutron star binary mergers during the current and next observational
runs at distances of 50 to 100\,Mpc, one will be in the position to
compare the merger events to each other. If such events involve
similar masses of primaries or secondaries with tidal deformabilities
that differ significantly, beyond the bounds of experimental error,
this would be an indication that different members of a twin or triplet
were involved in each event. Given the strong correlation between the
radii of stars and their tidal deformabilities, such observations
would be complementary to the possible inferences of different radii
of same-mass stars, as pointed out earlier~\cite{Alford2017}. This is
potentially detectable by the NICER mission~\cite{Gendreau2016}, which
is expected to constrain the radii of several x-ray emitting neutron stars
(in addition to the millisecond pulsar PSR J0030+0451) with
uncertainty about 10\%.

Looking ahead, the features and scenarios discussed in this work are
likely to be put to the test from various sides: (a) Observations of
massive pulsars in the future (for example with the Square Kilometer
Array) will put further constraints on the stiffness of the EOS at
high densities; (b) measurements of radii of neutron stars, such
as those already obtained by the NICER experiment, will provide
information on the EOS in the intermediate- to low- density range; (c)
future gravitational-wave observations of neutron star-neutron star
and neutron star-back hole binaries have the potential to further
narrow the range of admissible EOS. Increased statistics of
measurements will help to elucidate the possibility of first-order
phase transitions, as manifested by the occurrence of twins and
triplets of stars.

\section*{Acknowledgements}
We thank Steven Harris, Arus Harutyunyan, J\"urgen
Schaffner-Bielich and Fridolin Weber for discussions.
J.J.L. acknowledges the support of the Alexander von
Humboldt Foundation. A.S. acknowledges the support
of the DFG (Grant No. SE 1836/4-1). M.A. is supported
by the U.S. Department of Energy, Office of Science,
Office of Nuclear Physics under Award No.
DE-FG02-05ER41375. Partial support was provided
by the European COST Action ``PHAROS'' (CA16214) and
the State of Hesse LOEWE-Program in HIC for FAIR.
A.S. thanks David Blaschke and participants of the ECT$^*$
workshop ``The first compact star merger event --
Implications for nuclear and particle physics'' for
discussions. J.J.L. is also grateful for the hospitality
of Zhao-Qing Feng's group during his stay at the
South China University of Technology.

\bibliography{Hybridstars_refs}

\end{document}